\documentclass[aps,prl,reprint,groupedaddress]{revtex4-1}

\usepackage{color}
\usepackage[normalem]{ulem}

\usepackage{graphicx}

\newcommand{\Weber}{\mbox{\textit{We}}}
\newcommand{\Weberp}{\mbox{\textit{We$_p$}}}
\newcommand{\Reynolds}{\mbox{\textit{Re}}}

\newcommand{\Stokes}{\mbox{\textit{St}}}

\newcommand{\monodisperse}{{unimodal}}
\newcommand{\bidisperse}{{bimodal}}

\begin{document}

\title{Splashing onset in dense suspension droplets}

\author{Ivo R. Peters}
\email[]{irpeters@uchicago.edu}
\author{Qin Xu}
\author{Heinrich M. Jaeger}
\affiliation{James Franck Institute, The University of Chicago, Chicago, Illinois 60637, USA.}

\date{\today}

\begin{abstract}
We investigate the impact of droplets of dense suspensions onto a solid substrate. We show that a global hydrodynamic balance is unable to predict the splash onset and propose to replace it by an energy balance at the level of the particles in the suspension. We experimentally verify that the resulting, particle-based Weber number gives a reliable, particle size and density dependent splash onset criterion. We further show that the same argument also explains why in \bidisperse\ systems smaller particles are more likely to escape than larger ones.
\end{abstract}


\maketitle


%
%
Splashing of liquid droplets upon impact on a solid surface has been investigated for over a century~\cite{Worthington1876,Rein1993,Xu2005,Yarin2006,Xu2007a,Mandre2009,Tsai2011,Driscoll2011,Duchemin2011,Latka2012,Thoroddsen2012}. 
%
%
More recently, there has also been a growing interest in what happens to the spreading and splashing {if particles are added to the liquid~\cite{Nicolas2005,Luu2009,Guemas2012}}. On micron scales, ZrO$_2$ suspensions have been used in studies aiming to optimize ink-jet printing applications~\cite{Blazdell1995,Teng1997,Mott1999,Seerden2001,Lewis2006}, and on truly macroscopic scales there has been the development of 3D printers that dispense cement slurry~\cite{Pegna1997,Buswell2007}.  In all of these situations, an important concern is to prevent splashing, and particles from escaping, when droplets hit a surface. However, the question of when and why particles are ejected has remained unsettled, {and existing experimental studies mostly focus on dilute suspensions.}

%
%
{Current models for suspension drop impact associate the onset of splashing with the condition that $K=\Weber_d^{1/2}\Reynolds_d^{1/4}$  exceeds a critical value $K_0$, which has been the traditional criterion for pure liquid splashing on dry surfaces at atmospheric pressure in a regime independent of surface roughness~\cite{Mundo1995,Cossali1997,Yarin2006}. Here the 
Weber and Reynolds numbers are defined as $\Weber_d=\rho_lr_dU^2/\sigma$ and $\Reynolds_d=\rho_lr_dU/{\mu}$,
{with $r_d$ the droplet radius, $U$ the droplet impact velocity, and $\rho_l$, $\sigma$ and $\mu$ the liquid density, surface tension and dynamic viscosity, respectively.}

In these models, the addition of particles 
has  been captured by  replacing $\mu$ with an effective viscosity $\mu_e$ that increases with packing fraction 
~\cite{Eilers1941,Mooney1951,Krieger1959,Stickel2005,Nicolas2005,Bonnoit2012}.  This predicts that
a droplet of a pure liquid that would splash under certain conditions should not splash after adding enough particles. {To our knowledge, there exists no systematic study that confirms this prediction.} 
In fact, Nicolas observed~\cite{Nicolas2005} that adding particles, instead, lowered the splashing threshold $K_0$. 

%
%
\begin{figure*}
	\centering
	\includegraphics{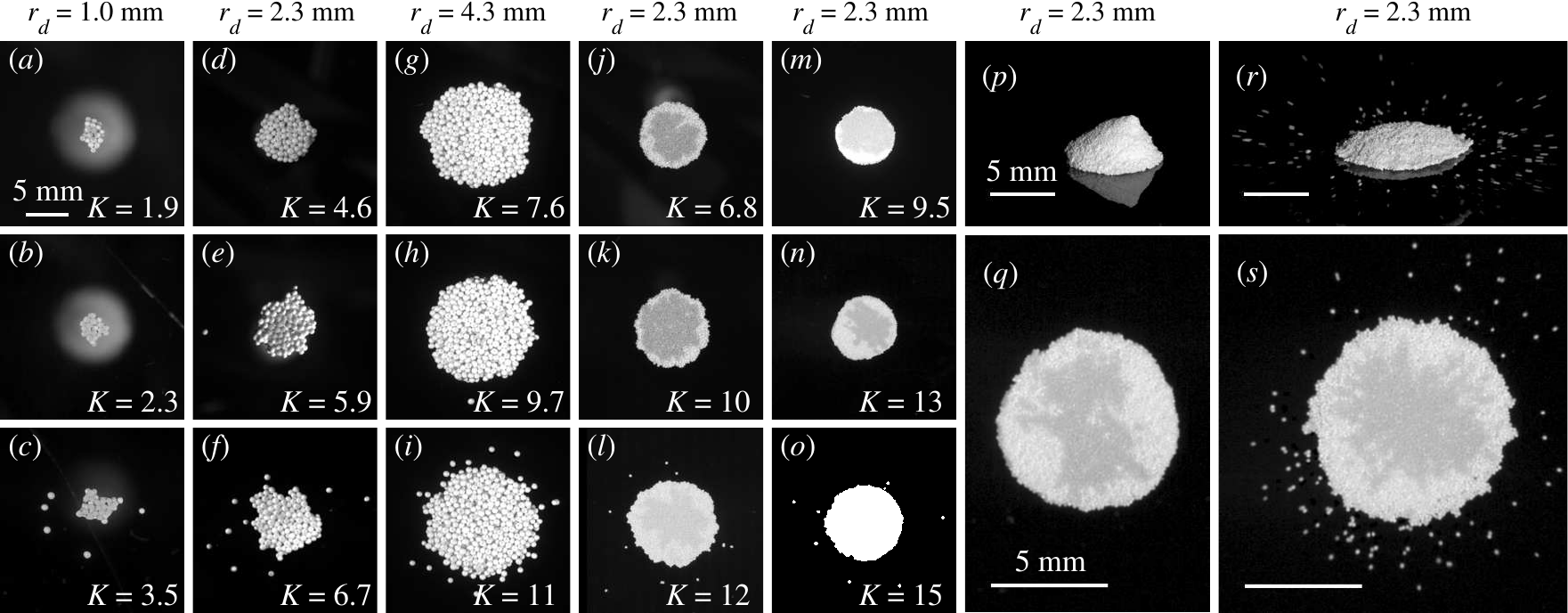}
	\caption{Still images right after impact of ZrO$_2$ particles in water onto a smooth glass plate, for different droplet radii {$r_d$} (see text above the images) and particle {radii}: ($a$-$i$) $r_p=362~\mathrm{\mu m}$, ($j$-$l$) $r_p=138~\mathrm{\mu m}$, ($m$-$s$) $r_p=78~\mathrm{\mu m}$. {The time between impact and ejection of the first particles ranges from $0.6$ to $3.5~\mathrm{ms}$.} {Images ($a$-$o$) are organized in vertical columns, with drop impact speed and therefore $K$ increasing from top to bottom to bracket the onset of splashing, defined as the ejection of individual particles.} {For the $K$ values listed in the top/middle/bottom rows we never/sometimes/always observed particle ejection.} $K=\Weber_d^{1/2}Re_d^{1/4}$, where we used the effective viscosity $\mu_e$ given by Krieger and Dougherty \cite{Krieger1959} in $Re_d$. The blurred background in ($a$) and ($b$) is the out-of-focus image of the syringe. Note that ($o$) has been thresholded and dilated in order visualize the ejected particles that would otherwise be invisible due to their small size. ($p$,$q$) side and bottom view of a droplet that does not splash ($\Weberp=12$), ($r$,$s$) side and bottom view of a splashing droplet ($\Weberp=26$). The scale bar in the images is $5~\mathrm{mm}$, images ($a$-$o$) all have the same scale.
	}
	\label{fig:splash_vs_nosplash}
\end{figure*}

To investigate 
the influence of added particles, we depart from the dilute limit described above, and instead 
focus here on the limit of 
dense {granular} suspensions with volume packing fractions {$\phi=0.62\pm0.03$}, where the discrepancy with the above droplet-scale splash onset criterion is most pronounced. 

In pure liquid droplets,  the size of the ejecta  depends on {either the destabilization of a thin liquid sheet~\cite{Allen1975,Thoroddsen1998,Thoroddsen2002,Xu2007a,Driscoll2010,Ruiter2010,Peters2013b} or, in the case of prompt splashing, on an instability at the moving contact line~\cite{Rioboo2002,Xu2007a,Latka2012}}. At splash onset in a suspension, on the other hand, the ejecta are the solid particles (see Fig.~\ref{fig:splash_vs_nosplash}), which implies a built-in length scale. This length scale was not taken into account in the energy balance leading to $K$, and not considered by {Refs.~\cite{Seerden2001,Nicolas2005,Lewis2006}}. 
{We will now evaluate the energy balance at the particle level, which we will then compare to our experiments.}

%
%
\begin{figure}
	\centering
	\includegraphics{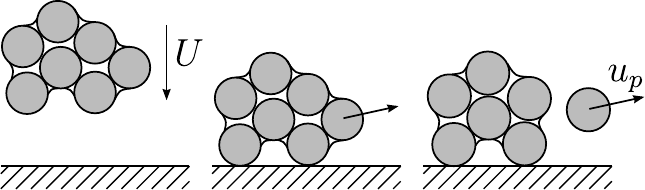}
	\caption{Sketch of the ejection mechanism upon impact.}
	\label{fig:ejectionCartoon}
\end{figure}

\emph{Energy balance.}--Surface tension keeps particles inside the drop because an escaping particle involves an increase of the surface energy, which scales with the particle surface area~\footnote{We neglect dewetting in this argument; typical dewetting speeds are of the order of $0.1~\mathrm{m/s}$~\cite{Blake1979,Winkels2011}. Partial dewetting would influence the actual surface energy, but keeps the scaling intact.}
\begin{equation}
	\label{eq:E_surf}
	{E_{surf} \sim 4\pi r_p^2\sigma},
\end{equation}
where $r_p$ is the particle radius. 
A particle can thus escape if it has enough kinetic energy, $E_{kin} = \frac{2}{3}\pi\rho_pr_p^3u_p^2$, with $\rho_p$ the specific density of a particle and $u_p$ its velocity, to overcome surface tension. The velocity $u_p$ is a result of collisions between neighboring particles (see Fig.~\ref{fig:ejectionCartoon}), 
which convert vertical into horizontal velocities. 
Based on momentum conservation we expect that the velocity {$u_p$} of a particle sitting on the outer surface of a droplet will be similar to the impact velocity of the drop $U$. The ratio of the kinetic and surface energy then is
\begin{equation}
	\label{eq:Weber}
	\frac{E_{kin}}{E_{surf}} {\sim} \frac{1}{6}\frac{\rho_p r_p U^2}{\sigma}
	\equiv\frac{1}{6}{\Weberp}.
\end{equation}
Here,
\Weberp\ is a particle-based Weber number. 
Fig.~\ref{fig:splash_vs_nosplash}($p$-$s$) shows {an example} {below and above the splashing onset}.


\emph{Experiments.}--We prepared suspensions of demineralized water with ZrO$_2$ {(Glenn Mills)} and soda-lime glass {(Mo-Sci)} particles inside a syringe. Particles were spherical, with standard deviations from their mean size of 5 to 8\% for the ZrO$_2$ and 15 to 20\% for the glass beads and densities $\rho_p=(3.9\pm 0.1)\cdot 10^3~\mathrm{kg/m^3}$ and $\rho_p=(2.53\pm 0.02)\cdot 10^3~\mathrm{kg/m^3}$, 
respectively. The volume fractions were determined by measuring the mass of a suspension drop, letting the water evaporate on a hot plate, and then measuring the mass of the dry particles. All packing fractions were between {59 and 65\%}. They are probably slightly overestimated due to finite sample size, because the amount of liquid depends on the shape of the menisci between the particles sitting at the surface of the droplet. In the current study we focus on changes to inviscid liquid splashing introduced by the particles and do not explore the role of additional viscous dissipation from the suspending liquid~\footnote{{Viscosity can play a role in either dissipating energy of escaping particles or changing the coefficient of restitution $e$, which is a function of the Stokes number $\Stokes=(2/9)\rho_pr_pU/\mu$ \cite{Gondret2002}. Stokes numbers for our experiments are in the range $\Stokes\approx100..2000$, which corresponds to $e\approx0.7..0.9$. We changed the viscosity by a factor two without observing a significant change in the splash onset. The splash onset is influenced if we change the viscosity more dramatically, i.e., by one order of magnitude, but this is outside the scope of our current study.}}.

Drops were formed by slowly pushing the suspension out of a cylindrical nozzle using a syringe pump. 
As gravity pulls the suspension down, a pinch-off will occur \cite{Miskin2012,Bertrand2012}, resulting in highly reproducible suspension drops. We varied $U$ by adjusting the release height, and $r_d$ by changing the nozzle size. {During extrusion of the suspension there is only minimal deformation of the droplet, 
producing a droplet radius equal to the nozzle radius~\footnote{{The droplets have smooth cylindrical shapes with an aspect ratio (height:width) of approximately 1:1. The orientation of this cylinder upon impact has no noticeable influence on the ejection of suspended particles.}}.} The experiments were recorded with a Phantom V12 high speed camera at frame rates of {$6.2-10~\mathrm{kHz}$} with a 105 mm Micro-Nikkor lens, resulting in a resolution of $20-50~\mathrm{\mu m/pixel}$. We used 
bottom views to observe and track particles ejected after impact. Typical bottom views are shown in Fig.~\ref{fig:splash_vs_nosplash}($a$-$o$). {We never observed the ejection of liquid droplets in our experiments.}

To determine the transition velocity $U^*$ above which particles are ejected from the droplet, we determine 
{the lower (upper) bound of $U^*$ (represented by the error bars in Fig. 3) at which we never (always) see ejected particles.}
Because we are able to distinguish individual particles, we define a non-splashing experiment when not a single particle leaves the suspension. A particle has left the suspension when there is no liquid bridge connecting it to the other particles. 
The experimental determination of $U^*$ for the {cases} of $r_p=78~\mathrm{\mu m}$ and $r_p=362~\mathrm{\mu m}$ ZrO$_2$ particles is shown in Fig.~\ref{fig:bidisperse}($a$), where $N_S$ is number of times we observe a splash and $N$ is the number of times we repeat one impact speed $U$ (typically 10 times).


%
%
\emph{Unimodal suspensions}--{Fig.~\ref{fig:splash_vs_nosplash} verifies that a global criterion, $K>K_0$,  does not capture the observed behavior. Within these examples there is about a factor 5 difference in $K_0$-values for droplets comprised of the same liquid and similar packing fraction of ZrO$_2$ beads. Note that in Fig.~\ref{fig:splash_vs_nosplash}($a$-$i$) the particles all have the same size.}
Fig.~\ref{fig:U_trans_vs_r} shows the influence of particle size on the splash transition. In all cases, the transition to splashing happens at the same value $\Weberp\approx14$. This is consistent with Eq.~(\ref{eq:Weber}), where {$\Weberp$} is the relevant parameter for the splash onset~\footnote{One might expect that the transition would occur at $\Weberp\approx6$, where $E_{kin}=E_{surf}$, but we stress that $u_p$ only scales with $U$, and is not necessarily identical to it. Additionally, an \textit{excess} of kinetic energy is needed in order to overcome the surface tension.}. Possible non-Newtonian effects or an effective viscosity seem to play no role here. Comparing the black, cyan, and magenta lines in Fig.~\ref{fig:bidisperse}($a$) --corresponding to the cluster of data points at the far right of Fig.~\ref{fig:U_trans_vs_r}-- we see that a factor of over four in $r_d$ has no significant influence on {$U^*$} -- in strong contrast with using {$r_d$} as {the} relevant length scale.

\begin{figure}
	\centering
	\includegraphics{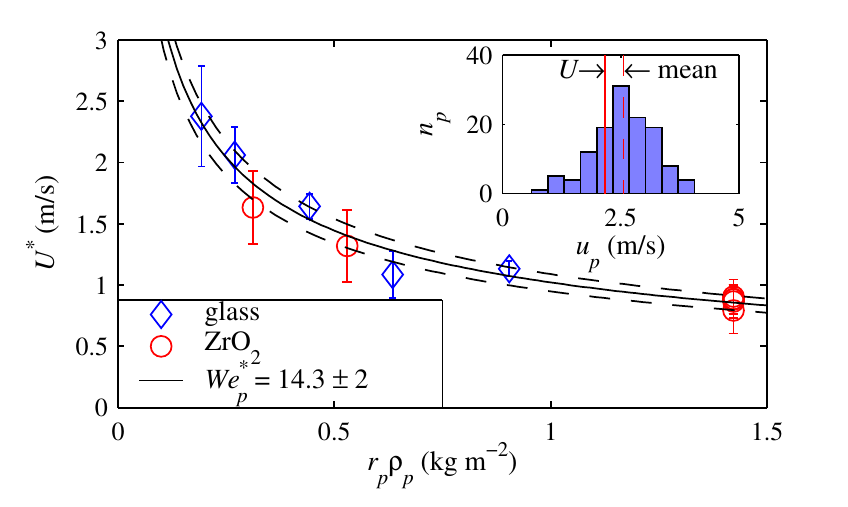}
	\caption{Splash onset velocity $U^*$ as function of the product of particle radius $r_p$ and particle density $\rho_p$. The cluster of 7 data points at the far right corresponds to 3 different drop sizes (see also Figs.~\ref{fig:splash_vs_nosplash},~\ref{fig:bidisperse}) and substrates with 4 different roughnesses {(see main text)}, all other data are for $r_d=2.3~\mathrm{mm}$ on a smooth substrate. The solid curve gives the onset Weber number $\Weberp^*$ calculated from a best fit to all experimental data. The dashed lines represent the upper and lower bounds for $\Weberp^*$, corresponding to one standard deviation. Inset: Histogram of the velocity $u_p$ of ejected particles at $\Weberp=20$, for $r_p=78~\mathrm{\mu m}$.}
	\label{fig:U_trans_vs_r}
\end{figure}

The inset in Fig.~\ref{fig:U_trans_vs_r} shows a typical velocity distribution for $78~\mathrm{\mu m}$ particles at $\Weberp=20$, where the mean velocity $\bar u_p$ of the particles after they are ejected from the suspension is slightly higher than the impact velocity $U$. For the larger ZrO$_2$ particles tested, $\bar u_p$ was smaller than $U$ and the ratio $\bar u_p/U$ decreases with particle size, but all measured mean velocities are in the range $0.5<\bar u_p/U<1.2$. Thus, $u_p$ of an ejected particle is always similar to the impact speed, confirming our estimate used in Eq.~(\ref{eq:Weber}).

%
%
\emph{Results on \bidisperse\ suspensions}--From Fig.~\ref{fig:bidisperse}($a$) it is clear that a suspension with particles of radius $362~\mathrm{\mu m}$ always splashes at {$U\gtrsim1.0~\mathrm{m/s}$}. On the other hand, a suspension with particles of radius $78~\mathrm{\mu m}$ never splashes at {$U\lesssim1.3~\mathrm{m/s}$}. So what happens if we make a \bidisperse\ suspension of these two particles types and impact at a speed between these two limits? We find that the splashing behavior is inverted: The larger particles remain inside, while the smaller particles get ejected. In Fig.~\ref{fig:bidisperse}($b$) we determine the splashing behavior for the \bidisperse\ suspension the same way as we did for \monodisperse\ suspensions in Fig.~\ref{fig:bidisperse}($a$). Clearly, the transition curves switch their position {when going from \monodisperse\ to \bidisperse .}

At first sight this result seems counterintuitive: surface tension should be more effective in keeping small particles inside the droplet than  large particles. To {qualitatively} explain this we take a closer look at how the particles obtain their velocity upon impact. 
We have argued before that
collisions between particles of the same size will result in similar velocities -- which explains why $u_p$ scales with $U$. Collisions between large and small particles can, however, result in much larger velocities for the small particles, which explains why the transition is at a lower impact speed. Conversely, small particles can only give  little velocity to large particles. The presence of small particles also reduces the chance for direct collisions between large particles, which explains the increased transition impact speed {for large particles in \bidisperse\ suspensions}.

\begin{figure}
	\centering
	\includegraphics{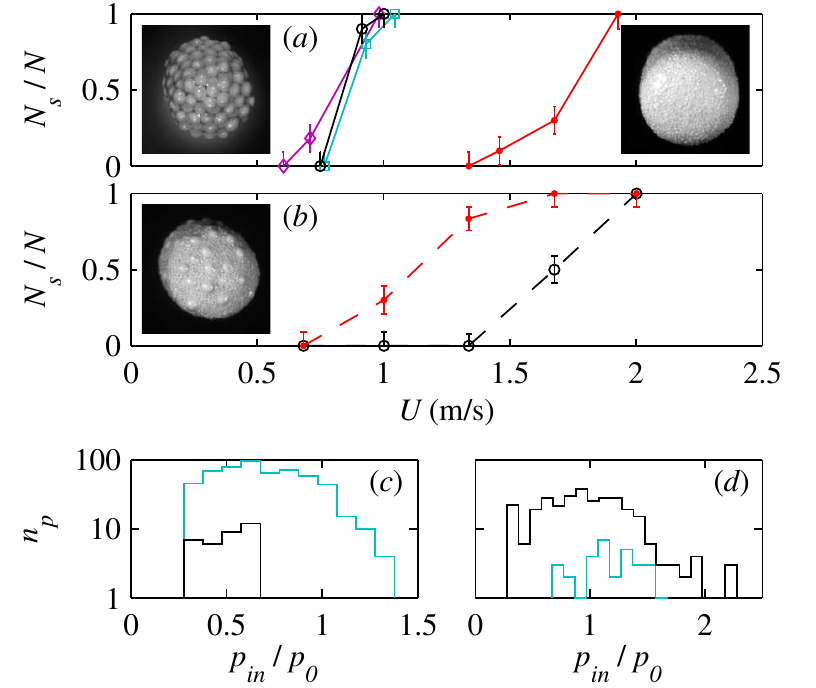}
	\caption{Comparison between \monodisperse\ and \bidisperse\ ZrO$_2$ suspensions. All are data for droplets made with nozzle radius $r_d=2.3~\mathrm{mm}$, except for data shown in magenta and cyan in ($a$) for which  $r_d=1.0~\mathrm{mm}$ and $r_d=4.3~\mathrm{mm}$, respectively. ($a$) Splash onset for \monodisperse\ suspensions of large ($r_p=362~\mathrm{\mu m}$, black, cyan, and magenta open symbols) and small ($r_p=78~\mathrm{\mu m}$, red dots) particles. ($b$) Splash onset for \bidisperse\ suspensions with volume ratio $\sim1:1$ of small to large particles. {Two distinct onsets exist: one where only small particles escape, and one where both small and large particles escape.} Data symbols and colors are as in ($a$). All insets: Examples of the  \monodisperse\ and \bidisperse\ suspension droplets {corresponding to the data presented in this figure}, imaged from below just before impact. ($c$) Histogram of relative momentum changes for large particles in \monodisperse\ (cyan) and \bidisperse\ (black) suspensions at $U=2~\mathrm{m/s}$. ($d$) Same as ($c$), but for small particles.
	}
	\label{fig:bidisperse}
\end{figure}

In order to extract the change in momentum due to the impact, we perform experiments for all three suspension configurations at an impact speed of $2~\mathrm{m/s}$ where we always find ejected particles: (i) \monodisperse\ suspensions with $r_p=362~\mathrm{\mu m}$, (ii) \monodisperse\ suspensions with $r_p=78~\mathrm{\mu m}$, and (iii) \bidisperse\ suspensions consisting of the same two different particles sizes. We calculate the gain in momentum due to the collisions as follows: 
We first 
determine the velocities of the ejected particles and calculate their kinetic energy. We know that during the ejection process the particle has transferred part of its kinetic energy to the surface energy given by Eq.~(\ref{eq:E_surf}). Adding this surface energy to the measured kinetic energy gives us the kinetic energy $E_{in}$ of the particle just \emph{before} it was ejected, which corresponds to the momentum $p_{in}=(\frac{8}{3}\pi\rho_pr_p^3E_{in})^{1/2}$. Comparing this to the vertical momentum of the particle before the moment of impact, $p_0=\frac{4}{3}\pi\rho_pr_p^3U$, gives us the relative change in momentum $p_{in}/p_0$ due to collisions between particles.

In Fig.~\ref{fig:bidisperse}($c\,$-$\,d$) we show $p_{in}/p_0$ for the three experiments mentioned above. Every impact speed was repeated 10 times and we count the number of ejected particles $n_p$ that are within a specific range of momenta. Fig.~\ref{fig:bidisperse}($c$) shows that for large particles going from \monodisperse\ suspensions (cyan) to \bidisperse\ suspensions (black), the probability of finding particles with a momentum in the order of $p_0$ decreases by at least an order of magnitude. Note that for the \bidisperse\ suspensions 
$p_{in}/p_0\lesssim0.75$. {For small particles on the other hand (Fig.~\ref{fig:bidisperse}($d$)), the probability of finding small particles with a momentum ratio $>1$ is increased by more than a factor 5 when changing from a \monodisperse\ to a \bidisperse\ suspension.}

Since collisions between particles are responsible for driving the onset of splashing, and less so the interaction of particles with the substrate, we expect that surface roughness will {play a much smaller role than for pure liquids}. To check this we performed experiments on substrates with roughness length scale $\ell$ ranging from {$\ell>r_p$ to $\ell\ll r_p$}. The results are included in Fig.~\ref{fig:U_trans_vs_r}, where in all cases there is no difference compared to the impact on a smooth surface. {The same independence of the splashing onset holds for experiments performed at a reduced ambient pressure $P\approx10~\mathrm{kPa}$.}


\emph{Conclusions and outlook.}--These experiments demonstrate that the relevant parameter to quantify the splash onset is a Weber number calculated at the particle level.  This is in contrast to earlier proposals for splash onset criteria in suspensions. Local interactions between particles at the edge of the droplet are responsible for the ejection of particles upon impact. This explains why the effective viscosity, which acts on a global droplet level, does not prevent the suspension from splashing. {Our observations give rise to the question at which packing fraction the global droplet description breaks down and when the dense limit, investigated here, takes over.} The same local interactions 
also drive the inversion of the splash onset for \bidisperse\ suspensions. {Since the momentum transfer in dense suspensions is collision dominated,  it might be possible to further tune the splash onset via particle characteristics 
such as
shape, restitution coefficient, or friction.  However, our findings already demonstrate that splash onset in dense suspensions  behaves qualitatively different  from predictions based on pure liquids. The typically used $K$ parameter does not delineate the splash onset properly.  Instead, a particle-based, critical Weber number $\Weberp^*$ describes the onset well.}

\begin{acknowledgments}
We thank C. Stevens for reading the manuscript and her experimental setup for testing the pressure dependence, and T. Witten, and W. Zhang for insightful discussions.  This work was supported by NSF through its MRSEC program (DMR-0820054) and by the US Army Research Office through grant number W911NF-12-1-0182.
\end{acknowledgments}

\bibliography{library}

\end{document}